\documentclass[12pt,english,draftcls, one column]{IEEEtran}
\usepackage[T1]{fontenc}
\usepackage[latin9]{inputenc}
\usepackage{float}
\usepackage{amsmath}
\usepackage{amsthm}
\usepackage{amssymb}
\usepackage{graphicx}
\usepackage{setspace}
\makeatletter

\floatstyle{ruled}
\newfloat{algorithm}{tbp}{loa}
\providecommand{\algorithmname}{Algorithm}
\floatname{algorithm}{\protect\algorithmname}

\theoremstyle{plain}

\theoremstyle{plain}

\ifCLASSINFOpdf
\else
\fi
\usepackage{babel}

\makeatother

\usepackage{babel}
\providecommand{\propositionname}{Proposition}
\providecommand{\theoremname}{Theorem}

\date{}
\begin{document}

\title{Robust Baseband Compression Against Congestion in Packet-Based Fronthaul Networks Using Multiple Description Coding}

\author{Seok-Hwan Park, \textit{Member}, \textit{IEEE}, Osvaldo Simeone,
\textit{Fellow, IEEE}, \\ and Shlomo Shamai (Shitz),\textit{ Fellow,
IEEE} \thanks{
S.-H. Park was supported by the National Research Foundation of Korea (NRF) grant funded by the Korea government {[}NRF-2018R1D1A1B07040322{]}. The work of O. Simeone was partially supported by the U.S. NSF through grant 1525629 and by the European Research Council (ERC) under the European Union\textquoteright s Horizon 2020 research and innovation programme (grant agreement No 725731). The work of S. Shamai has been supported by the European Union\textquoteright s Horizon 2020 research and innovation programme, grant agreement no. 694630.

S.-H. Park is with the Division of Electronic Engineering, Chonbuk
National University, Jeonju 54896, Korea (email: seokhwan@jbnu.ac.kr).

O. Simeone is with the Department of Informatics, King\textquoteright s
College London, London, UK (email: osvaldo.simeone@kcl.ac.uk).

S. Shamai (Shitz) is with the Department of Electrical Engineering,
Technion, Haifa, 32000, Israel (email: sshlomo@ee.technion.ac.il).}}

\maketitle
\begin{abstract}
In modern implementations of Cloud Radio Access Network (C-RAN), the fronthaul transport network will often be packet-based and it will have a multi-hop architecture built with general-purpose switches using network function virtualization (NFV) and software-defined networking (SDN). This paper studies the joint design of uplink radio and fronthaul transmission strategies for a C-RAN with a packet-based fronthaul network. To make an efficient use of multiple routes that carry fronthaul packets from remote radio heads (RRHs) to cloud, as an alternative to more conventional packet-based multi-route reception or coding, a multiple description coding (MDC) strategy is introduced that operates directly at the level of baseband signals. MDC ensures an improved quality of the signal received at the cloud in conditions of low network congestion, i.e., when more fronthaul packets are received within a tolerated deadline. The advantages of the proposed MDC approach as compared to the traditional path diversity scheme are validated via extensive numerical results.\end{abstract}


\theoremstyle{theorem}
\newtheorem{theorem}{Theorem}
\theoremstyle{proposition}
\newtheorem{proposition}{Proposition}
\theoremstyle{lemma}
\newtheorem{lemma}{Lemma}
\theoremstyle{corollary}
\newtheorem{corollary}{Corollary}
\theoremstyle{definition}
\newtheorem{definition}{Definition}
\theoremstyle{remark}
\newtheorem{remark}{Remark}

\section{Introduction} \label{sec:intro}
In a Cloud Radio Access Network (C-RAN) architecture, a~cloud unit, or~baseband processing unit (BBU), carries out baseband signal processing on behalf of a number of radio units, or~remote radio heads (RRHs), that are connected to the cloud through an interface referred to as fronthaul links~\cite{Checko}. The~C-RAN technology is recognized as one of the dominant architectural solutions for future wireless networks due to the promised reduction in capital and operational expenditures and the capability of large-scale interference management~\cite{Simeone:JCN16}. A~major challenge of C-RAN deployment is that high-rate baseband in-phase and quadrature (IQ) samples need to be carried on the fronthaul links of limited data rate. The~design of signal processing strategies, including fronthaul compression techniques, for~C-RAN was widely studied in the literature~\cite{Park:TVT13,Park:TSP13,Zhou:TSP16,Patil:TCOM18}.

\begin{figure}[H]
\centering\includegraphics[width=13.5cm,height=5.5cm]{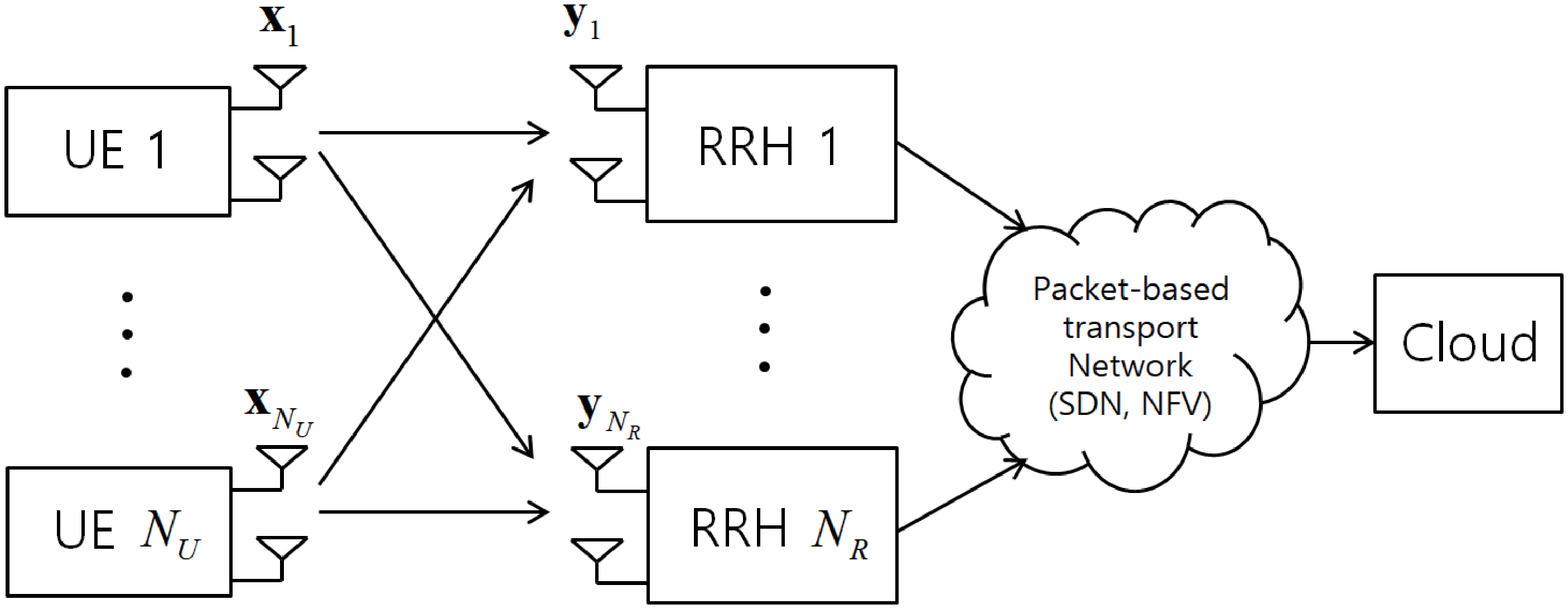}

\caption{\label{fig:system-model}Illustration of the uplink of a Cloud Radio Access Network (C-RAN) with a packet-based fronthaul transport~network.}
\end{figure}

The mentioned works~\cite{Park:TVT13,Park:TSP13,Zhou:TSP16,Patil:TCOM18} and references therein assume a conventional fronthaul topology, whereby there are dedicated point-to-point fronthaul links from the cloud to each RRH as in Common Public Radio Interface (CPRI) specification~\cite{CPRI}. However, in~modern implementations of C-RAN, as~illustrated in Figure~\ref{fig:system-model}, the~fronthaul transport network will often be packet-based and it will have a multi-hop architecture built with general-purpose switches using network function virtualization (NFV) and software-defined networking (SDN) \cite{Oliva:CM16,eCPRI}. Packet-based fronthaul network can leverage the wide deployment of Ethernet infrastructure~\cite{Ethernet}.

Packet-based multi-hop networks are subject to congestion and packet losses. The~traditional path diversity approach repeats the same packet on the multiple routes  in order to mitigate these issues~\cite{Alasti:TIT01,Mountaser}. This approach can successfully reduce the packet loss probability at the cost of increasing the overhead in the fronthaul network. A~limitation of these traditional schemes is that, when multiple packets arrive at the cloud within the tolerated delay, the~signal quality utilized for channel decoding at the cloud is the same as if a single packet is received. To~make a more efficient use of the multiple routes, in~this paper, we propose a multiple description coding (MDC) scheme that operates directly on the baseband signals. Thanks to MDC, a~better distortion level is obtained as more packets arrive at the cloud within the deadline. We refer to~\cite{Goyal:SPM01} for an overview and for a discussion on applications of MDC. In~addition, the~work~\cite{Benammar} proposed the use of MDC to improve the achievable rate of a multicast cognitive interference~channel.

Since, thanks to MDC, the~signal quality varies depending on the number of packets arriving at the cloud, we propose that user equipments (UEs) leverage the broadcast approach in order to enable the adaptation of the transmission rate to the effective received signal-to-noise ratio (SNR) \cite{Cover,Shamai-Steiner}.
The broadcast approach defines a variable-to-fixed channel code~\cite{Verdu} that enables the achievable rate to adapt to the channel state when the latter is known only at the receiving end.
The broadcast approach splits the message of each UE into multiple submessages that are encoded independently, and~transmitted as a superposition of the encoded signals. With~the proposed MDC-based solution, based on the packets received  within a given deadline, the~cloud performs  successive interference cancellation (SIC) decoding of the UEs' submessages with a given order so that the achievable rate can be adapted to the number of delivered packets. Therefore, the~number of received packets determines the quality of the channel state known only at the receiver.
Related methods were introduced in~\cite{Park-et-al:TVT14} and~\cite{Karasik}, where broadcast coding with layered  compression~\cite{Ng} was applied to the uplink of C-RAN systems with distributed channel state information~\cite{Park-et-al:TVT14} and with uncertain fronthaul capacity~\cite{Karasik}.

\begin{figure}[H]
\centering\includegraphics[width=14.5cm,height=5.4cm]{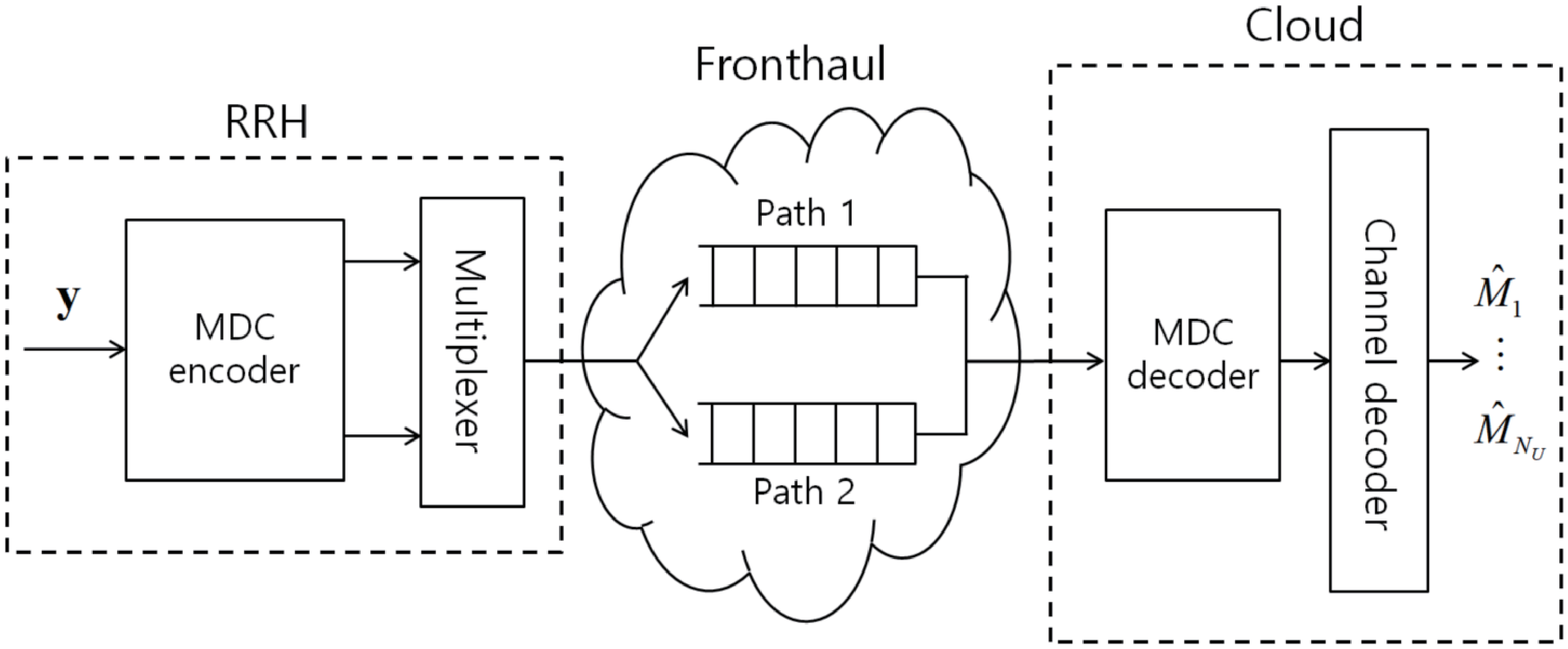}

\caption{\label{fig:MDC}Illustration of the multiple description coding (MDC)
and packet-based transport network.}
\end{figure}

More specifically, in~this work, we study joint radio and fronthaul transmission for the uplink of a C-RAN with a packet-based fronthaul network. In~the system, the~uplink received baseband signal of each RRH is quantized and compressed producing a bit stream. The~output bits are then packetized and transmitted on the fronthaul network. Following a standard approach to increase robustness to network losses and random delays (see, e.g.,~\cite{Alasti:TIT01,Mountaser}), we assume that the packets are sent over multiple paths towards the cloud as seen in Figure~\ref{fig:MDC}. This can be done by using either conventional packet-based duplication~\cite{Alasti:TIT01,Mountaser} or the proposed MDC approach. The~packets may be lost due to network delays or congestion when they are not received within a tolerable fronthaul delay dependent on the application.
Based on the packets that have arrived within the delay, the~cloud carries out decompression and channel~decoding.

The rest of the paper is organized as follows. In~Section~\ref{sec:System-Model}, we describe the system model for the uplink of a C-RAN with packet-based fronthaul network. In~Section~\ref{sec:Robust-Compression-Based-on-MDC-2paths}, we present the proposed MDC scheme which operates in a combination with the broadcast coding. The~optimization of the proposed scheme is discussed in Section~\ref{sec:Problem-Definition-and}, and~the advantages of the proposed scheme are validated with extensive numerical results in Section~\ref{sec:numerical}. We discuss extension to general cases in Section~\ref{sec:extension}, and~the paper is concluded in Section~\ref{sec:conclusion}.

We summarize some notations used throughout the paper as follows. The~mutual information between random variables $X$ and $Y$ conditioned on $Z$ is denoted as $I(X;Y|Z)$, and~$h(X)$ denotes the differential entropy of $X$. We define $\mathcal{CN}(\mbox{\boldmath${\mu}$}, \mathbf{\Sigma})$ as the circularly symmetric complex Gaussian distribution with mean $\mbox{\boldmath${\mu}$}$ and covariance $\mathbf{\Sigma}$. The~expectation, trace, determinant and Hermitian transpose operations are denoted by $\mathtt{E}(\cdot)$, $\text{tr}(\cdot)$, $\det(\cdot)$ and $(\cdot)^H$, respectively, and~$\mathbb{C}^{M\times N}$ represents the set of all $M\times N$ complex matrices. We denote as $\mathbf{I}_N$ an identity matrix of size $N$, and~$\otimes$ represents the Kronecker product. $\mathbf{A} \succeq \mathbf{0}$ indicates that the matrix $\mathbf{A}$ is positive~semidefinite.

\section{System Model\label{sec:System-Model}}

We consider the uplink of a C-RAN in which $N_{U}$ UEs communicate with a cloud unit through $N_R$ RRHs. To~emphasize the main idea, we first focus on the case of $N_R=1$ and discuss extension to a general number of RRHs in Section~\ref{sec:extension}.
Also, for~convenience, we define the set $\mathcal{N}_{U}\triangleq\{1,\ldots,N_{U}\}$
of UEs, and~denote the numbers of antennas of UE $k$ and of the RRH
by $n_{U,k}$ and $n_{R}$, respectively. The~key novel aspect as compared to the prior work reviewed above is the assumption of packet-based fronthaul connecting between RRH and~cloud.

\subsection{Uplink Wireless~Channel}

Each UE $k$ encodes its message to be decoded at the cloud and obtains
an encoded baseband signal $\mathbf{x}_{k}\sim\mathcal{CN}(\mathbf{0},\mathbf{\Sigma}_{\mathbf{x}_{k}})\in\mathbb{C}^{n_{U,k}\times1}$
which is transmitted on the uplink channel toward the RRH. Assuming
flat-fading channel, the~signal $\mathbf{y}\in\mathbb{C}^{n_{R}\times1}$
received by the RRH is given as
\begin{equation}
\mathbf{y}=\sum_{k\in\mathcal{N}_{U}}\mathbf{H}_{k}\mathbf{x}_{k}+\mathbf{z}=\mathbf{H}\mathbf{x}+\mathbf{z},\label{eq:received-signal-RRH-i}
\end{equation}
where $\mathbf{H}_{k}\in\mathbb{C}^{n_{R}\times n_{U,k}}$ is the
channel transfer matrix from UE $k$ to the RRH, $\mathbf{z}\sim\mathcal{CN}(\mathbf{0},\mathbf{\Sigma}_{\mathbf{z}})$
is the additive noise vector, $\mathbf{H}=[\mathbf{H}_{1}\,\cdots\,\mathbf{H}_{N_{U}}]$
is the channel matrix from all the UEs to the RRH, and~$\mathbf{x}=[\mathbf{x}_{1}^{H}\,\cdots\,\mathbf{x}_{N_{U}}^{H}]^{H}\sim\mathcal{CN}(\mathbf{0},\mathbf{\Sigma}_{\mathbf{x}})$
is the signal transmitted by all the UEs with $\mathbf{\Sigma}_{\mathbf{x}}=\mathrm{diag}(\{\mathbf{\Sigma}_{\mathbf{x}_{k}}\}_{k\in\mathcal{N}_{U}})$.
We define the covariance matrix $\mathbf{\Sigma}_{\mathbf{y}}=\mathbf{H}\mathbf{\Sigma}_{\mathbf{x}}\mathbf{H}^{H}+\mathbf{\Sigma}_{\mathbf{z}}$
of $\mathbf{y}$.

The RRH quantizes and compresses the received signal $\mathbf{y}$
producing a number of packets. As~detailed next, these packets are sent to the cloud on a packet-based fronthaul network,
and the cloud jointly decodes the messages sent by the UEs based on
the signals received within some maximum allowed fronthaul~delay.

\subsection{Packet-Based Fronthaul Transport~Network}

As discussed in Section~\ref{sec:intro}, in~modern implementations of C-RAN, the~fronthaul transport network
is expected to be packet-based and to have a multi-hop architecture
built with general-purpose switches using NFV and SDN~\cite{Oliva:CM16,eCPRI}.
As a result, upon~compression, the~received signals need to be packetized,
and the packets to be transmitted on the fronthaul network to the
cloud. Packets may be lost due to network delays or congestion when
they are not received within a tolerated fronthaul delay dependent on the~application.

\begin{figure}[H]
\centering\includegraphics[width=13.5cm,height=5.5cm]{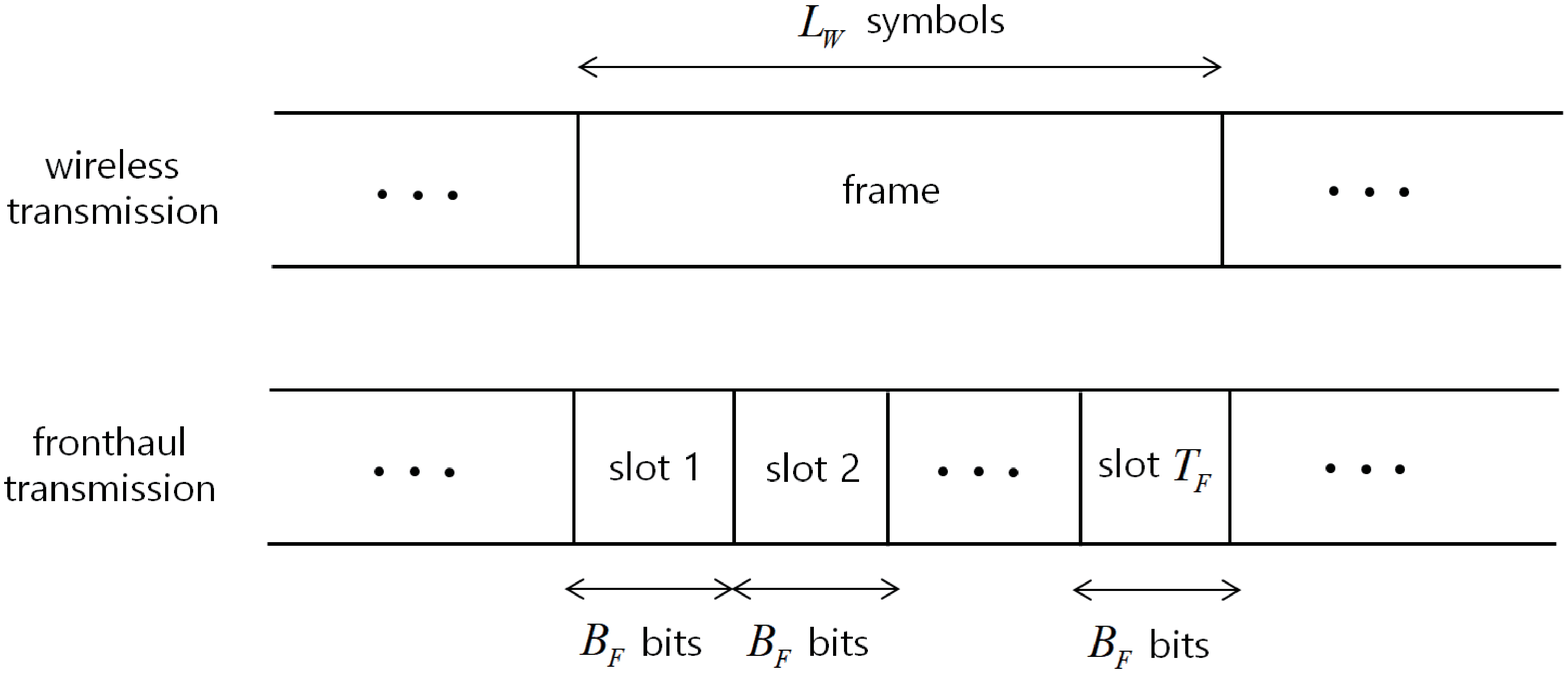}

\caption{\label{fig:timescale}Illustration of wireless frame and fronthaul slotted~transmission.}
\end{figure}

A standard approach to increase robustness to network losses and random
delays is to send packets over multiple paths towards the destination
(see, e.g.,~\cite{Alasti:TIT01,Mountaser}). As~seen in Figure~\ref{fig:MDC}, following~\cite{Alasti:TIT01}, we model transmission on each such path, or~route, as~a queue. Furthermore, as~seen in Figure~\ref{fig:timescale}, transmission on the fronthaul transport network is slotted, with~each slot carrying a payload of $B_F$ bits. The~duration of each
wireless frame, of~$L_W$ symbols, encompasses $T_{F}$ fronthaul slots. Due to congestion, each fronthaul packet sent by the RRH on route $j$ takes a geometrically
distributed number of time slots to be delivered. Accordingly, transmission is successful independently in each slot with probability $1-\epsilon_{F,j}$ on the $j$th~route.

\unskip

\section{Robust Compression Based on Multiple Description Coding\label{sec:Robust-Compression-Based-on-MDC-2paths}}

In this section, we propose a robust compression technique based on MDC, which, in~combination with broadcast coding, enables the achievable rate to be adapted to the number of packets collected by the cloud, and~hence to the current network congestion level.
To highlight the idea, we assume that the RRH has available two paths
to the cloud. Extensions will be discussed in Section~\ref{sec:extension}.
The~traditional path diversity approach repeats the
same packet on the two routes~\cite{Alasti:TIT01}. More sophisticated forms of packet-based encoding, such as erasure coding studied in~\cite{Mountaser}, are not applicable to the case of two paths. Accordingly, if~one
or two packets are received by the fronthaul deadline of $T_{F}$ slots, the~signal is decompressed and decoding is carried out at the cloud. Note
that, if~both packets are received, the~signal quality is the same
as if one packet is received. In~contrast, we propose to adopt MDC as seen in Figure~\ref{fig:MDC}.
With MDC, if~one packet is received by the deadline $T_{F}$, we obtain a certain
distortion level, while we obtain a better distortion level if both
packets are received (\cite[Ch. 14]{ElGamal-Kim}).

In the MDC approach, the~RRH first quantizes and compresses the received
signal $\mathbf{y}$ to produce quantized signals $\hat{\mathbf{y}}_{0}$,
$\hat{\mathbf{y}}_{1}$ and $\hat{\mathbf{y}}_{2}$. Packets $\hat{\mathbf{y}}_{1}$ and $\hat{\mathbf{y}}_{2}$ are sent on two separate paths to the cloud, with~$\hat{\mathbf{y}}_{l}$ sent on route $l$. By~the properties of MDC, if~only a single
packet $l\in\{1,2\}$ arrives at the cloud within deadline $T_{F}$,
the MDC decoder can recover the quantized signal $\hat{\mathbf{y}}_{l}$,
while the signal $\hat{\mathbf{y}}_{0}$ can be recovered if the both
packets are received in~time.

Denote as $R_{F}$ the number of bits per symbol
used to represent the signal for each of the quantized packets $\hat{\mathbf{y}}_{1}$ and $\hat{\mathbf{y}}_{2}$. We refer to $R_{F}$ as the compression output rate.
As shown in \cite[Ch. 14]{ElGamal-Kim}, the~rate $R_{F}$ should satisfy the conditions
\begin{align}
R_{F} & \geq I\left(\mathbf{y};\hat{\mathbf{y}}_{1}\right),\label{eq:constraint-MDC-1}\\
R_{F} & \geq I\left(\mathbf{y};\hat{\mathbf{y}}_{2}\right),\label{eq:constraint-MDC-2}\\
\text{and }\,\,2R_{F} & \geq I\left(\mathbf{y};\{\hat{\mathbf{y}}_{l}\}_{l\in\{0,1,2\}}\right)+I\left(\hat{\mathbf{y}}_{1};\hat{\mathbf{y}}_{2}\right).\label{eq:constraint-MDC-12}
\end{align}
To evaluate (\ref{eq:constraint-MDC-1})--(\ref{eq:constraint-MDC-12}), as~in, e.g.,~\cite{Park:TVT13,Park:TSP13,Zhou:TSP16,Patil:TCOM18},
we assume standard Gaussian quantization codebooks, so that the quantized
signals can be modeled as
\begin{equation}
\hat{\mathbf{y}}_{l}=\mathbf{y}+\mathbf{q}_{l},\label{eq:quantized-signals-MDC}
\end{equation}
for $l\in\{0,1,2\}$, where the quantization noise $\mathbf{q}_{l}$
is independent of the signal $\mathbf{y}$ and distributed as $\mathbf{q}_{l}\sim\mathcal{CN}(\mathbf{0},\mathbf{\Omega})$ for $l\in\{1,2\}$ and $\mathbf{q}_{0}\sim\mathcal{CN}(\mathbf{0},\mathbf{\Omega}_0)$.
The right-hand sides (RHSs) of (\ref{eq:constraint-MDC-1})--(\ref{eq:constraint-MDC-12})
can hence be written as
\begin{align}
g_{l}\left(\mathbf{\Omega}, \mathbf{\Omega}_0\right)= & I\left(\mathbf{y};\hat{\mathbf{y}}_{l}\right)\label{eq:compression-rate-individual}\\
= & \log_{2}\det\left(\mathbf{\Sigma}_{\mathbf{y}}+\mathbf{\Omega}\right)-\log_{2}\det\left(\mathbf{\Omega}\right),\,\,l\in\{1,2\},\nonumber \\
\text{and }\,\,g_{\mathrm{sum}}\left(\mathbf{\Omega}, \mathbf{\Omega}_0\right)= & I\left(\mathbf{y};\{\hat{\mathbf{y}}_{l}\}_{l\in\{0,1,2\}}\right)+I\left(\hat{\mathbf{y}}_{1};\hat{\mathbf{y}}_{2}\right)\label{eq:compression-rate-sum}\\
= & h\left(\mathbf{y}\right)+h\left(\{\hat{\mathbf{y}}_{l}\}_{l\in\{0,1,2\}}\right)-h\left(\mathbf{y},\{\hat{\mathbf{y}}_{l}\}_{l\in\{0,1,2\}}\right)\nonumber \\
 & +h\left(\hat{\mathbf{y}}_{1}\right)+h\left(\hat{\mathbf{y}}_{2}\right)-h\left(\hat{\mathbf{y}}_{1},\hat{\mathbf{y}}_{2}\right)\nonumber \\
= & \log_{2}\det\left(\mathbf{\Sigma}_{\mathbf{y}}\right)+\log_{2}\det\left(\mathbf{A}_{3}\mathbf{\Sigma}_{\mathbf{y}}\mathbf{A}_{3}^{H}+\bar{\mathbf{\Omega}}\right)\nonumber \\
 & -\log_{2}\det\left(\mathbf{A}_{4}\mathbf{\Sigma}_{\mathbf{y}}\mathbf{A}_{4}^{H}+\mathrm{diag}(\mathbf{0}_{n_{R}},\bar{\mathbf{\Omega}})\right)\nonumber \\
 & +2\log_{2}\det\left(\mathbf{\Sigma}_{\mathbf{y}}+\mathbf{\Omega}\right) -\log_{2}\det\left(\mathbf{A}_{2}\mathbf{\Sigma}_{\mathbf{y}}\mathbf{A}_{2}^{H}+ \mathbf{I}_2 \otimes \mathbf{\Omega} \right), \nonumber
\end{align}
where we have defined the notations
$\bar{\mathbf{\Omega}}=\mathrm{diag}(\mathbf{\Omega}_0, \mathbf{I}_2 \otimes \mathbf{\Omega})$ and
$\mathbf{A}_{m}=\mathbf{1}_{m}\otimes\mathbf{I}_{n_{R}}$ with $\mathbf{1}_{m}\in\mathbb{C}^{m\times1}$
denoting a column vector of all ones.

We now discuss the derivation of the probability that a packet $l$
is delivered to the cloud within the given deadline $T_F$.
The number $N_{F}$ of fronthaul packets that need to be delivered
within the time $T_{F}$ to the cloud for the $l$th description is
given as
\begin{equation}
N_{F}=\left\lceil \frac{L_{W}R_{F}}{B_{F}}\right\rceil ,\label{eq:Number-of-packets}
\end{equation}
since $L_{W}R_{F}$ is the number of bits per description and $B_{F}$
is the number of available bits per frame. Note that $N_F$ increases with the compression output rate $R_F$ and decreases with the size of the fronthaul packet $B_F$. Then, the~probability that
description $l\in\{1,2\}$ sent on route $l$ is received
at the cloud within the deadline $T_{F}$ is given as
\begin{equation}
P_{l}^{c}(T_{F})=\mathrm{Pr}\left[\sum_{m=1}^{N_{F}}T_{l,m}\leq T_{F}\right],\label{eq:probability-success-arrival}
\end{equation}
where $\{T_{l,m}\}_{m=1}^{N_{F}}$ are independent and geometrically
distributed random variables with parameter $1-\epsilon_{F,l}$ such
that the sum $\sum_{m=1}^{N_{F}}T_{l,m}$ is a negative binomial random
variable with parameters $1-\epsilon_{F,l}$ and $N_{F}$ (\cite[Ch. 3]{LGarcia}).
Therefore, the~probability (\ref{eq:probability-success-arrival})
can be written as
\begin{equation}
P_{l}^{c}(T_{F})=1-I_{\epsilon_{F,l}}\left(T_{F}-N_{F}+1,\,N_{F}\right),
\end{equation}
where $I_{x}(a,b)$ is the regularized incomplete beta function defined
as
\begin{equation}
I_{x}(a,b)=\frac{B(x;a,b)}{B(1;a,b)},\label{eq:regularized-incomplete-beta-function}
\end{equation}
with $B(x;a,b)=\int_{0}^{x}t^{a-1}(1-t)^{b-1}dt$. For~simplicity
of notation, we also define the probabilities $P_{\emptyset}^{c}(T_{F})=(1-P_{1}^{c}(T_{F}))(1-P_{2}^{c}(T_{F}))$
and $P_{\text{all}}^{c}(T_{F})=P_{1}^{c}(T_{F})P_{2}^{c}(T_{F})$ that no or
both descriptions arrive at the cloud within the~deadline.

Define as $M\in\{0,1,2\}$ the number of descriptions
that arrive at the cloud within the given deadline $T_{F}$. The~probability
distribution $p_{M}(m) = \Pr\left[M=m\right]$
can then be written as
\begin{equation}
p_{M}(m)=\begin{cases}
P_{\emptyset}^{c}(T_{F}), & m=0\\
\sum_{l=1}^{2}P_{l}^{c}(T_{F})\left(1-P_{\bar{l}}^{c}(T_{F})\right), & m=1\\
P_{\text{all}}^{c}(T_{F}), & m=2
\end{cases},\label{eq:probability-of-index-li}
\end{equation}
with the notation $\bar{1}=2$ and $\bar{2}=1$.

\subsection{Broadcast Coding\label{subsec:Broadcast-Coding}}

With MDC, the~quality of the information available at the cloud
for decoding the transmitted signals $\{\mathbf{x}_{k}\}_{k\in\mathcal{N}_{U}}$
is determined by the number $M$ of descriptions that arrive at the
cloud. Since the state $M\in\{0,1,2\}$ is not known to the UEs, the~rate cannot be a priori adapted by the UEs depending on the congestion level. To~handle this issue,  we
propose that each UE $k$ adopts a broadcast coding strategy~\cite{Cover,Shamai-Steiner,Verdu,Park-et-al:TVT14}
as
\begin{equation}
\mathbf{x}_{k}=\mathbf{x}_{k,1}+\mathbf{x}_{k,2},\label{eq:broadcast-coding-each-UE}
\end{equation}
where the signals $\mathbf{x}_{k,1}$ and $\mathbf{x}_{k,2}$ encode
independent messages of UE $k$, and~the decoder at the cloud is required
to reliably recover only the signals $\{\mathbf{x}_{k,j}\}_{k\in\mathcal{N}_{U}}$
with $j\leq m$ when $M=m$ descriptions arrive at the cloud. We denote
the rate of the signal $\mathbf{x}_{k,m}$ as $R_{k,m}$ for $k\in\mathcal{N}_{U}$
and $m\in\{1,2\}$. We make the standard assumption that the $j$th signal
$\mathbf{x}_{k,j}$ of each UE $k$ is distributed as $\mathbf{x}_{k,j}\sim\mathcal{CN}(\mathbf{0},P_{k,j}\mathbf{I}_{n_{U,k}})$, where the powers $P_{k,j}$ need to satisfy the power constraint $P_{k,1}+P_{k,2}=P$. Under~the described assumption, the~covariance
matrix $\mathbf{\Sigma}_{\mathbf{x}}$ of all the transmitted signals
$\mathbf{x}$ is given as $\mathbf{\Sigma}_{\mathbf{x}}=P\mathbf{I}_{n_{U}}$
with $n_{U}=\sum_{k\in\mathcal{N}_{U}}n_{U,k}$.

The signal $\mathbf{r}_{m}$ collected at the cloud
when $M=m$ descriptions have arrived at the cloud is given as
\begin{align}
\mathbf{r}_{m}= & \begin{cases}
\mathbf{0}, & m=0\\
\hat{\mathbf{y}}_{1}, & m=1\\
\hat{\mathbf{y}}_{0}, & m=2
\end{cases}.\label{eq:quantized-signal-collected}
\end{align}
For the case of $m=1$, the~cloud receives $\mathbf{r}_{1}=\hat{\mathbf{y}}_{1}$
or $\mathbf{r}_{1}=\hat{\mathbf{y}}_{2}$. In~(\ref{eq:quantized-signal-collected}),
we set $\mathbf{r}_{1}=\hat{\mathbf{y}}_{1}$ without loss of generality,
since $\hat{\mathbf{y}}_{1}$
and $\hat{\mathbf{y}}_{2}$ are statistically~equivalent.

When no description arrives at the cloud (i.e., $M=0$), the~cloud
has no information received from the RRH, and~none of the signals
$\{\mathbf{x}_{k,0},\mathbf{x}_{k,1}\}_{k\in\mathcal{N}_{U}}$ can
be decoded by the cloud. When only a single description arrives at
the cloud ($M=1$), the~cloud jointly decodes the first-layer signals
$\{\mathbf{x}_{k,1}\}_{k\in\mathcal{N}_{U}}$ based on the received
quantized signal $\mathbf{r}_{1}$. Therefore, the~achievable sum-rate
$R_{\Sigma,1}=\sum_{k\in\mathcal{N}_{U}}R_{k,1}$ of the first-layer signals
is given as
\begin{align}
R_{\Sigma,1} & =f_{1}\left(\mathbf{P},\mathbf{\Omega}, \mathbf{\Omega}_0\right)=I\left(\bar{\mathbf{x}}_{1};\mathbf{r}_{1}\right)\label{eq:sum-rate-layer-1}\\
 & =\log_{2}\det\left(\mathbf{H}\mathbf{\Sigma}_{\mathbf{x}}\mathbf{H}^{H}+\mathbf{\Sigma}_{\mathbf{z}}+\mathbf{\Omega}\right)-\log_{2}\det\left(\mathbf{H}\bar{\mathbf{P}}_{2}\mathbf{H}^{H}+\mathbf{\Sigma}_{\mathbf{z}}+\mathbf{\Omega}\right),\nonumber
\end{align}
where we have defined the vector $\bar{\mathbf{x}}_{m}=[\mathbf{x}_{1,m}^{H}\,\cdots\,\mathbf{x}_{N_{U},m}^{H}]^{H}\sim\mathcal{CN}(\mathbf{0},\bar{\mathbf{P}}_{m})$
that stacks the layer-$m$ signals of all the UEs, and~the notations
$\mathbf{P}=\{P_{k,j}\}_{k\in\mathcal{N}_{U},j\in\{1,2\}}$ and $\bar{\mathbf{P}}_{m}=\mathrm{diag}(\{P_{k,m}\}_{k\in\mathcal{N}_{U}})$.

If both descriptions arrive at the cloud (i.e., $M=2$),
the cloud first jointly decodes the first-layer signals $\{\mathbf{x}_{k,1}\}_{k\in\mathcal{N}_{U}}$
from the recovered quantized signal $\mathbf{r}_{2}$, and~cancels
the impact of the decoded signals from $\mathbf{r}_{2}$, i.e.,~$\tilde{\mathbf{r}}_{2}\leftarrow\mathbf{r}_{2}-\sum_{k\in\mathcal{N}_{U}}\mathbf{H}_{k}\mathbf{x}_{k,1}$.
Then, the~cloud decodes the second-layer signals $\{\mathbf{x}_{k,2}\}_{k\in\mathcal{N}_{U}}$
based on $\tilde{\mathbf{r}}_{2}$. Thus, the~achievable sum-rate
$R_{\Sigma,2}=\sum_{k\in\mathcal{N}_{U}}R_{k,2}$ of the second-layer signals
is given as
\begin{align}
R_{\Sigma,2} & =f_{2}\left(\mathbf{P},\mathbf{\Omega}, \mathbf{\Omega}_0\right)=I\left(\bar{\mathbf{x}}_{2};\mathbf{r}_{2}|\bar{\mathbf{x}}_{1}\right)\label{eq:sum-rate-each-layer}\\
 & =\log_{2}\det\left(\mathbf{H}\bar{\mathbf{P}}_{2}\mathbf{H}^{H}+\mathbf{\Sigma}_{\mathbf{z}}+\mathbf{\Omega}_{0}\right)-\log_{2}\det\left(\mathbf{\Sigma}_{\mathbf{z}}+\mathbf{\Omega}_{0}\right).\nonumber
\end{align}

\begin{figure}[H]
\centering  \includegraphics[width=18.0cm,height=13.0cm]{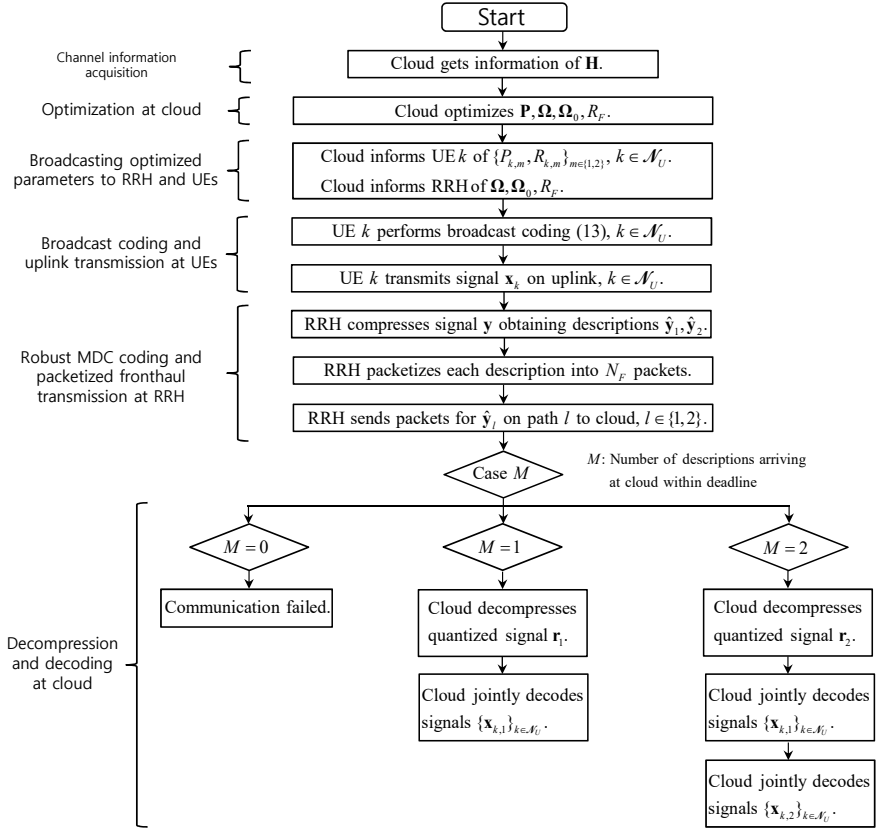}

\caption{\label{fig:flow-chart}A flowchart that illustrates the operations of the proposed uplink system based on broadcast coding and multiple description coding (MDC).}
\end{figure}

In summary, the~whole system operates as follows. The~cloud first obtains the channel state information and optimizes the variables related to broadcast coding and MDC coding. The~optimization will be discussed in Section~\ref{sec:Problem-Definition-and}. After~the optimization algorithm is finished, the~cloud informs the UEs and the RRH of the optimized variables. The~UEs perform broadcast coding and uplink transmission, and~the RRH compresses the received signal obtaining two descriptions which are packetized and sent on fronthaul paths to the cloud. Based on the received packets, the~cloud performs MDC decoding of the quantized signals and SIC decoding of the UEs' messages.
We provide a flowchart that illustrates the described operations of the proposed system in Figure~\ref{fig:flow-chart}.

\section{Problem Definition and Optimization\label{sec:Problem-Definition-and}}

For fixed instantaneous channel states $\{\mathbf{H}_{k}\}_{k\in\mathcal{N}_{U}}$,
we aim at jointly optimizing the compression output rate $R_{F}$,
the power allocation variables $\mathbf{P}$ and the quantization noise
covariance matrices $\{\mathbf{\Omega}, \mathbf{\Omega}_0\}$ with the goal of maximizing
the expected sum-rate denoted as $\bar{R}_{\Sigma}$. Here the expectation
is taken with respect to the random variables $\{T_{l,m}\}_{l\in\{1,2\},m\in\mathcal{N}_{F}}$
with $\mathcal{N}_{F}=\{1,2,\ldots,N_{F}\}$, which depend on the current congestion level of the packet network. The~expected sum-rate
$\bar{R}_{\Sigma}$ is hence given as
\begin{align}
\bar{R}_{\Sigma} & =p_{M}(1)R_{\Sigma,1}+p_{M}(2)\left(R_{\Sigma,1}+R_{\Sigma,2}\right)\label{eq:expected-sum-rate-definition}\\
 & =\bar{p}_{M}(1)R_{\Sigma,1}+\bar{p}_{M}(2)R_{\Sigma,2},\nonumber
\end{align}
with the notations $\bar{p}_{M}(1)=p_{M}(1)+p_{M}(2)$ and $\bar{p}_{M}(2)=p_{M}(2)$.
The expected sum-rate $\bar{R}_{\Sigma}$ can be expressed as a function
of $R_{F}$, $\mathbf{P}$ and $\{\mathbf{\Omega},\mathbf{\Omega}_0\}$:
\begin{align}
\bar{R}_{\Sigma} & =f_{\Sigma}\left(R_{F},\mathbf{P},\mathbf{\Omega},\mathbf{\Omega}_0\right)\label{eq:expected-sum-rate}\\
 & =\bar{p}_{M}(1)f_{1}\left(\mathbf{P},\mathbf{\Omega},\mathbf{\Omega}_0\right)+\bar{p}_{M}(2)f_{2}\left(\mathbf{P},\mathbf{\Omega},\mathbf{\Omega}_0\right).\nonumber
\end{align}

 We note that increasing the compression output rate $R_{F}$ has
conflicting effects on the expected sum-rate $\bar{R}_{\Sigma}$. On~the one hand, the~probability of timely reception of all fronthaul packets
decreases with $R_{F}$ due to the increased number $N_{F}$ of packets
in (\ref{eq:Number-of-packets}). On~the other hand, once the packets have arrived at the cloud, a~better sum-rate can be achieved with larger $R_{F}$,
since the quantization noise signals have smaller~powers.

The problem mentioned above can be stated as\begin{subequations}\label{eq:problem-original}
\begin{align}
\underset{R_{F},\mathbf{P},\mathbf{\Omega},\mathbf{\Omega}_0}{\mathrm{maximize}} & \,\,\,f_{\Sigma}\left(R_{F},\mathbf{P},\mathbf{\Omega},\mathbf{\Omega}_0\right)\label{eq:problem-original-objective}\\
\mathrm{s.t.}\,\,\, & R_{F}\geq g_{1}\left(\mathbf{\Omega},\mathbf{\Omega}_0\right),\label{eq:problem-original-compression-rate-individual}\\
 & 2R_{F}\geq g_{\mathrm{sum}}\left(\mathbf{\Omega},\mathbf{\Omega}_0\right),\label{eq:problem-original-compression-rate-sum}\\
 & \mathbf{\Omega}\succeq\mathbf{0},\,\,\mathbf{\Omega}_{0}\succeq\mathbf{0},\label{eq:problem-original-psd}\\
 & P_{k,1}+P_{k,2}=P,\,\,k\in\mathcal{N}_{U},\label{eq:problem-original-power-sum}\\
 & P_{k,1}\geq0,\,\,P_{k,2}\geq0,\,\,k\in\mathcal{N}_{U}.\label{eq:problem-original-power-nonnegative}
\end{align}
\end{subequations}
To tackle the problem (\ref{eq:problem-original}),
we first note that, if~we fix the compression output
rate variable $R_{F}$, the~problem becomes a difference-of-convex (DC)
problem as in~\cite{Park-et-al:SPL}. Therefore, we can find an efficient solution
by adopting the concave convex procedure (CCCP) approach (see, e.g.,
\cite{Tao-et-al:TWC,Park-et-al:TWC}). The~detailed algorithm
that tackles (\ref{eq:problem-original}) with the CCCP approach is
described in Algorithm~\ref{algorithm1}, where we have defined the functions $\tilde{f}_{\Sigma}(R_{F},\mathbf{P},\mathbf{\Omega},\mathbf{\Omega}_0,\mathbf{P}^{(t)},\mathbf{\Omega}^{(t)},\mathbf{\Omega}_0^{(t)})$,
$\tilde{g}_{1}(\mathbf{\Omega},\mathbf{\Omega}_0,\mathbf{\Omega}^{(t)},\mathbf{\Omega}_0^{(t)})$ and $\tilde{g}_{\mathrm{sum}}(\mathbf{\Omega},\mathbf{\Omega}_0,\mathbf{\Omega}^{(t)},\mathbf{\Omega}_0^{(t)})$
as
\begin{align*}
\tilde{f}_{\Sigma}\left(R_{F},\mathbf{P},\mathbf{\Omega},\mathbf{\Omega}_0,\mathbf{P}^{(t)},\mathbf{\Omega}^{(t)},\mathbf{\Omega}_0^{(t)}\right)= & \bar{p}_{M}(1)\left(\begin{array}{c}
\log_{2}\det\left(\mathbf{H}\mathbf{\Sigma}_{\mathbf{x}}\mathbf{H}^{H}+\mathbf{\Sigma}_{\mathbf{z}}+\mathbf{\Omega}\right)\\
-\phi\left(\begin{array}{c}
\mathbf{H}\bar{\mathbf{P}}_{2}\mathbf{H}^{H}+\mathbf{\Sigma}_{\mathbf{z}}+\mathbf{\Omega},\\
\mathbf{H}\bar{\mathbf{P}}_{2}^{(t)}\mathbf{H}^{H}+\mathbf{\Sigma}_{\mathbf{z}}+\mathbf{\Omega}^{(t)}
\end{array}\right)
\end{array}\right)\\
 & +\bar{p}_{M}(2)\left(\begin{array}{c}
\log_{2}\det\left(\mathbf{H}\bar{\mathbf{P}}_{2}\mathbf{H}^{H}+\mathbf{\Sigma}_{\mathbf{z}}+\mathbf{\Omega}_{0}\right)\\
-\phi\left(\mathbf{\Sigma}_{\mathbf{z}}+\mathbf{\Omega}_{0},\mathbf{\Sigma}_{\mathbf{z}}+\mathbf{\Omega}_{0}^{(t)}\right)
\end{array}\right),\\
\tilde{g}_{1}\left(\mathbf{\Omega},\mathbf{\Omega}_0,\mathbf{\Omega}^{(t)},\mathbf{\Omega}_0^{(t)}\right)= & \phi\left(\mathbf{\Sigma}_{\mathbf{y}}+\mathbf{\Omega},\mathbf{\Sigma}_{\mathbf{y}}+\mathbf{\Omega}^{(t)}\right)-\log_{2}\det\left(\mathbf{\Omega}\right),\\
\text{and }\,\,\tilde{g}_{\mathrm{sum}}\left(\mathbf{\Omega},\mathbf{\Omega}_0,\mathbf{\Omega}^{(t)},\mathbf{\Omega}_0^{(t)}\right)= & \log_{2}\det\left(\mathbf{\Sigma}_{\mathbf{y}}\right)+\phi\left(\mathbf{A}_{3}\mathbf{\Sigma}_{\mathbf{y}}\mathbf{A}_{3}^{H}+\bar{\mathbf{\Omega}},\mathbf{A}_{3}\mathbf{\Sigma}_{\mathbf{y}}\mathbf{A}_{3}^{H}+\bar{\mathbf{\Omega}}^{(t)}\right)\\
 & -\log_{2}\det\left(\mathbf{A}_{4}\mathbf{\Sigma}_{\mathbf{y}}\mathbf{A}_{4}^{H}+\mathrm{diag}(\mathbf{0}_{n_{R}},\bar{\mathbf{\Omega}})\right)\\
 & +2\phi\left(\mathbf{\Sigma}_{\mathbf{y}}+\mathbf{\Omega},\mathbf{\Sigma}_{\mathbf{y}}+\mathbf{\Omega}^{(t)}\right) 
 -\log_{2}\det\left(\mathbf{A}_{2}\mathbf{\Sigma}_{\mathbf{y}}\mathbf{A}_{2}^{H}+ \mathbf{I}_2 \otimes \mathbf{\Omega} \right),
\end{align*}
with the function $\phi\left(\mathbf{A},\mathbf{B}\right)$ defined
as
\begin{align*}
\phi\left(\mathbf{A},\mathbf{B}\right) & =\log_{2}\det(\mathbf{B})+\frac{1}{\ln2}\mathrm{tr}\left(\mathbf{B}^{-1}(\mathbf{A}-\mathbf{B})\right).
\end{align*}

\begin{algorithm}[H]
\caption{CCCP algorithm for problem (\ref{eq:problem-original}) for fixed
$R_{F}$}\label{algorithm1}

\textbf{1.} Initialize the variables $\mathbf{P}^{(1)}$, $\mathbf{\Omega}^{(1)},\mathbf{\Omega}_0^{(1)}$
to arbitrary matrices that satisfy the constraints (\ref{eq:problem-original-compression-rate-individual}),
(\ref{eq:problem-original-compression-rate-sum}) and (\ref{eq:problem-original-psd}),
and set $t\leftarrow1$.

\textbf{2.} Update the variables $\mathbf{P}^{(t+1)},\mathbf{\Omega}^{(t+1)},\mathbf{\Omega}_0^{(t+1)}$
as a solution of the convex problem:\begin{subequations}\label{eq:problem-convexified}
\begin{align}
\underset{\mathbf{P},\mathbf{\Omega},\mathbf{\Omega}_0}{\mathrm{maximize}} & \,\,\,\tilde{f}_{\Sigma}\left(R_{F},\mathbf{P},\mathbf{\Omega},\mathbf{\Omega}_0,\mathbf{P}^{(t)},\mathbf{\Omega}^{(t)},\mathbf{\Omega}_0^{(t)}\right)\label{eq:problem-original-objective-1}\\
\mathrm{s.t.}\,\,\, & R_{F}\geq\tilde{g}_{1}\left(\mathbf{\Omega},\mathbf{\Omega}_0,\mathbf{\Omega}^{(t)},\mathbf{\Omega}_0^{(t)}\right),\label{eq:problem-original-compression-rate-individual-1}\\
 & 2R_{F}\geq\tilde{g}_{\mathrm{sum}}\left(\mathbf{\Omega},\mathbf{\Omega}_0,\mathbf{\Omega}^{(t)},\mathbf{\Omega}_0^{(t)}\right),\label{eq:problem-original-compression-rate-sum-1}\\
 & \mathbf{\Omega}\succeq\mathbf{0},\,\,\mathbf{\Omega}_0\succeq\mathbf{0},\label{eq:problem-original-psd-1}\\
 & P_{1}+P_{2}=P,\\
 & P_{1}\geq0,\,\,P_{2}\geq0.
\end{align}
\end{subequations}

\textbf{3.} Stop if a convergence criterion is satisfied. Otherwise,
set $t\leftarrow t+1$ and go back to Step 2.
\end{algorithm}

We have discussed the optimization of the power allocation variables
$\mathbf{P}$ and the quantization noise covariance matrices $\{\mathbf{\Omega},\mathbf{\Omega}_0\}$
for fixed compression output rate $R_{F}$. For~the optimization of
$R_{F}$, we propose to perform a 1-dimensional discrete search over
$R_{F}\in\mathcal{R}=\{\Delta_{R_{F}},2\Delta_{R_{F}},\ldots,N_{F,\max}\Delta_{R_{F}}\}$
with $\Delta_{R_{F}}=B_{F}/L_{W}$ and $N_{F,\max}=T_{F}+1$. Here
we have excluded the values $\tau\Delta_{R_{F}}$ with non-integer
$\tau$ from the search space $\mathcal{R}$. This does not cause
a loss of optimality, since we can increase the compression output
rate, hence improving the compression fidelity to $\left\lceil \tau\right\rceil \Delta_{R_{F}}$
without increasing the number $N_{F}$ of packets in (\ref{eq:Number-of-packets})
that needs to be delivered to the~cloud.

\subsection{Optimization of Traditional Path-Diversity Scheme\label{sub:Optimization-of-Traditional-Path-Diversity}}

In this subsection, we discuss the optimization of the traditional
path-diversity (PD) scheme, in~which the RRH repeats to send the same
packet on the available two routes~\cite{Alasti:TIT01}. Accordingly, the~RRH produces only a single quantized signal $\hat{\mathbf{y}}=\mathbf{y}+\mathbf{q}$,
where the quantization noise $\mathbf{q}$ is independent of $\mathbf{y}$
and distributed as $\mathbf{q}\sim\mathcal{CN}(\mathbf{0},\mathbf{\Omega})$
under the assumption of standard Gaussian quantization codebooks.
Denoting as $R_{F}$ the compression output rate for the quantized
signal $\hat{\mathbf{y}}$, the~rate $R_{F}$ should satisfy the condition
\begin{align}
R_{F}\geq & g\left(\mathbf{\Omega}\right)=I\left(\mathbf{y};\hat{\mathbf{y}}\right)\label{eq:constraint-PD}\\
 & =\log_{2}\det\left(\mathbf{\Sigma}_{\mathbf{y}}+\mathbf{\Omega}\right)-\log_{2}\det\left(\mathbf{\Omega}\right).\nonumber
\end{align}

To evaluate the achievable sum-rate, we define the binary variable
$D\in\{0,1\}$, which takes 1 if at least one packet arrives at the
cloud, and~0 otherwise. The~probability distribution of $D$ can be
written~as
\begin{equation}
\Pr\left[D=d\right]=\begin{cases}
P_{\emptyset}^{c}(T_{F}), & d=0\\
1-P_{\emptyset}^{c}(T_{F}), & d=1
\end{cases}.\label{eq:probability-variable-c}
\end{equation}

If both packets sent on two routes are lost (i.e., $D=0$), the~cloud
cannot decode the signals sent by the UEs. If~the cloud receives at
least one packet ($D=1$), the~cloud can perform decoding of the signals
$\mathbf{x}$ based on the received quantized signal $\hat{\mathbf{y}}$,
and the achievable sum-rate can be written as
\begin{align}
R_{\Sigma} & =f_{\Sigma}(\mathbf{\Omega})=I\left(\mathbf{x};\hat{\mathbf{y}}\right)\label{eq:sum-rate-PD}\\
 & =\log_{2}\det\left(\mathbf{H}\mathbf{\Sigma}_{\mathbf{x}}\mathbf{H}^{H}+\mathbf{\Sigma}_{\mathbf{z}}+\mathbf{\Omega}\right)-\log_{2}\det\left(\mathbf{\Sigma}_{\mathbf{z}}+\mathbf{\Omega}\right).\nonumber
\end{align}
The expected sum-rate $\bar{R}_{\Sigma}$ can be expressed as
\begin{align}
\bar{R}_{\Sigma} & =f_{\Sigma}\left(R_{F},\mathbf{\Omega}\right)=\Pr\left[D=1\right]f_{\Sigma}(\mathbf{\Omega}).\label{eq:expected-sum-rate-PD}
\end{align}

The problem of maximizing the expected sum-rate $\bar{R}_{\Sigma}$
with the traditional PD scheme can hence be stated as\begin{subequations}\label{eq:problem-original-PD}
\begin{align}
\underset{R_{F},\mathbf{\Omega}}{\mathrm{maximize}} & \,\,\,f_{\Sigma}\left(R_{F},\mathbf{\Omega}\right)\label{eq:problem-original-objective-2}\\
\mathrm{s.t.}\,\,\, & R_{F}\geq g\left(\mathbf{\Omega}\right),\label{eq:problem-original-compression-rate-PD}\\
 & \mathbf{\Omega}\succeq\mathbf{0}.\label{eq:problem-original-psd-PD}
\end{align}
\end{subequations}We can tackle the problem (\ref{eq:problem-original-PD})
in a similar approach to that proposed for addressing (\ref{eq:problem-original}).

\section{Numerical Results \label{sec:numerical}}

In this section, we provide numerical results that validate the advantages
of the proposed robust baseband compression technique based on MDC coding scheme.
We consider a system bandwidth of 100 MHz and assume that each wireless
frame consists of $L_{W}=5000$ channel uses. We also assume that
each fronthaul packet has $B_{F}=6000$ bits (i.e., 750 bytes) which
corresponds to a half of the maximum payload size per frame defined
in Ethernet~\cite{Ethernet}. Denoting as $C_{F}$ the fronthaul capacity
in bit/s, each fronthaul packet has the duration of $B_{F}/C_{F}$.
If we define the maximum tolerable delay on fronthaul network as $T_{\max}$
s, the~deadline $T_{F}$ in packet duration is given as $T_{F}=\left\lfloor T_{\max}/(B_{F}/C_{F})\right\rfloor $.
In the simulation, we set $T_{\max}=1$ ms. For~simplicity, we assume
that all paths have the same error probability $\epsilon_{F,l}=\epsilon_{F}$
for all $l\in\{1,2\}$. Regarding the channel statistics, we assume
that the positions of the UEs and the RRH are uniformly distributed
within a circular area of radius $100$ m. The~elements of the channel
matrix $\mathbf{H}_{k}$ are independent and identically distributed
(i.i.d.) as $\mathcal{CN}(0,\rho_{k})$. Here the path-loss $\rho_{k}$
is modeled as $\rho_{k}=1/(1+(d_{k}/d_{0})^{3})$, where $d_{k}$
represents the distance between the RRH  and UE $k$, and~$d_{0}$
is the reference distance set to $d_{0}=30$ m.
We set the noise covariance to $\mathbf{\Sigma}_{\mathbf{z}} = N_0 \mathbf{I}_{n_R}$, and~the SNR is defined as $P/N_0$.

\subsection{Fixed Compression Output Rate $R_{F}$}

\begin{figure}[H]
\centering\includegraphics[width=11cm,height=9cm,keepaspectratio]{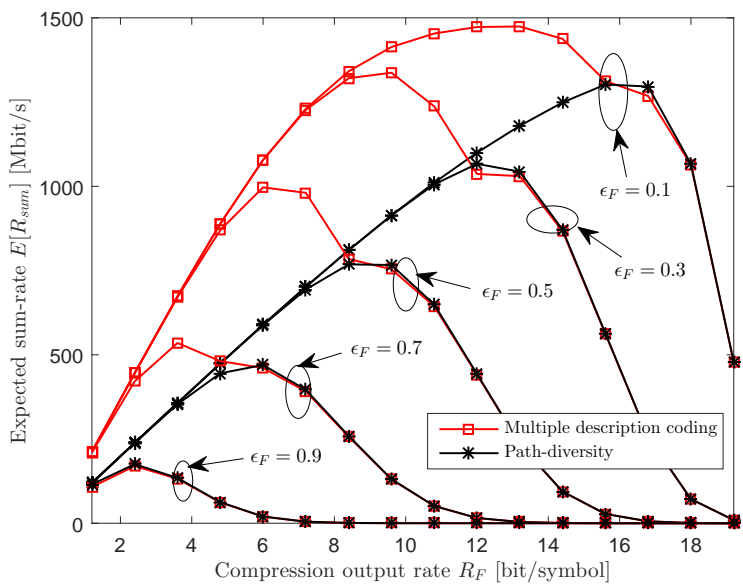}\caption{{\scriptsize{}\label{fig:graph-Rsum-vs-RF-variousMu}}{\footnotesize{}Expected
sum-rate $\mathtt{E}[R_{\mathrm{sum}}]$ versus the compression output
rate $R_{F}$ for various values of $\epsilon_{F}\in\{0.1,0.3,0.5,0.7,0.9\}$
($N_{U}=2$, $n_{R}=2$, $n_{U,k}=1$, $C_{F}=100$ Mbit/s and 25
dB signal-to-noise ratio (SNR)).}}
\end{figure}

We first evaluate the expected
sum-rate performance $\mathtt{E}[R_{\mathrm{sum}}]$ when only the power allocation variables $\mathbf{P}$ and the quantization noise covariance matrices $\mathbf{\Omega}$ are optimized
according to Algorithm 1 for fixed compression output rate $R_{F}$.
In Figure~\ref{fig:graph-Rsum-vs-RF-variousMu}, we plot the expected
sum-rate $\mathtt{E}[R_{\mathrm{sum}}]$ versus the compression output
rate $R_{F}$ for various values of path error probability $\epsilon_{F}$
with $N_{U}=2$, $n_{R}=2$, $n_{U,k}=1$, $C_{F}=100$ Mbit/s and
25 dB SNR. We observe that, for~both the MDC and PD schemes, the~optimal
compression output rate $R_{F}$ increases as the fronthaul error
probability $\epsilon_{F}$ decreases. This suggests that, with~smaller
$\epsilon_{F}$, the~packet networks become more reliable and hence
more packets can be reliably delivered to the cloud within the deadline. Furthermore,
the figure shows that, with~MDC, it is optimal to choose a lower compression
output rate with respect to PD. This is because, as~the fronthaul
quality improves in terms of the error probability $\epsilon_{F}$,
the PD scheme can only increase the sum-rate by increasing the quality,
or the compression output rate $R_{F}$, of~each individual description,
since it cannot benefit from reception of both descriptions. In~contrast,
the MDC scheme can operate at a lower $R_{F}$, since the quality
of the compressed signal is improved by reception of both descriptions.
Receiving both descriptions tends to be more likely if the compression
output rate is lower and hence the number of fronthaul packets per
frame is reduced.

\begin{figure}[H]
\centering\includegraphics[width=11cm,height=9cm,keepaspectratio]{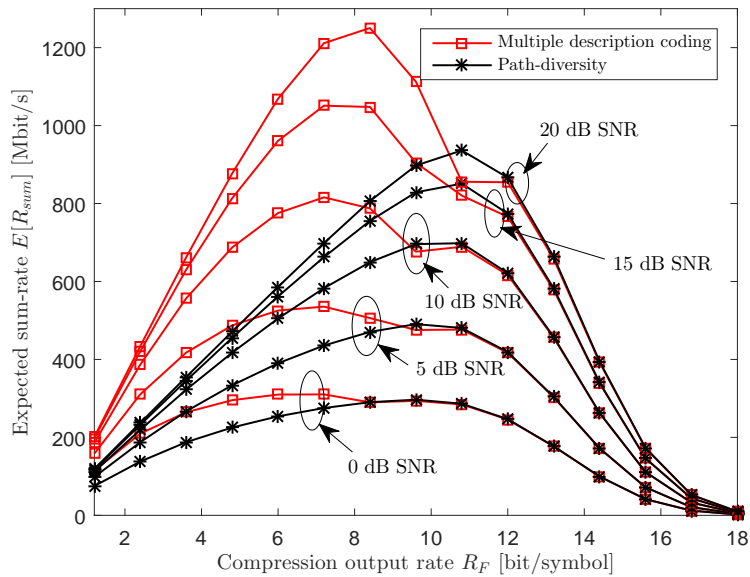}\caption{{\scriptsize{}\label{fig:graph-Rsum-vs-RF-variousSNR}}{\footnotesize{}Expected
sum-rate $\mathtt{E}[R_{\mathrm{sum}}]$ versus the compression output
rate $R_{F}$ for various SNR levels ($\epsilon_{F}=0.4$, $N_{U}=3$,
$n_{R}=3$, $n_{U,k}=1$ and $C_{F}=100$ Mbit/s).}}
\end{figure}

In Figure~\ref{fig:graph-Rsum-vs-RF-variousSNR}, we depict the expected
sum-rate $\mathtt{E}[R_{\mathrm{sum}}]$ versus the compression output
rate $R_{F}$ for various SNR levels with $\epsilon_{F}=0.4$, $N_{U}=3$,
$n_{R}=3$, $n_{U,k}=1$ and $C_{F}=100$ Mbit/s. The~figure shows
that, as~the SNR increases, the~optimal compression output rate $R_{F}$
slightly increases for both MDC and PD. This is because, while the
SNR level does not affect the reliability of the packet fronthaul
network, it is desirable for the RRH to report better descriptions
of the uplink received signals to the cloud when the received signals
carry more information on the UEs'~messages.

\begin{figure}[H]
\centering\includegraphics[width=11cm,height=9cm,keepaspectratio]{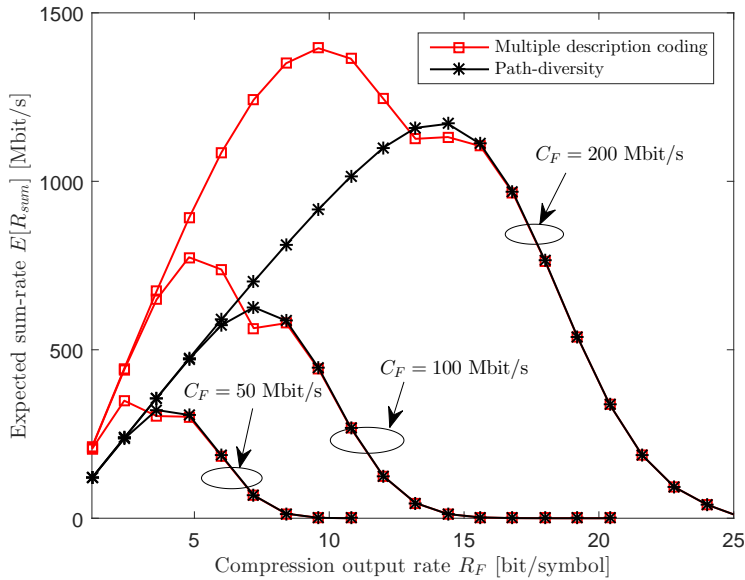}\caption{{\scriptsize{}\label{fig:graph-Rsum-vs-RF-variousCF}}{\footnotesize{}Expected
sum-rate $\mathtt{E}[R_{\mathrm{sum}}]$ versus the compression output
rate $R_{F}$ for various fronthaul capacity $C_{F}$ ($\epsilon_{F}=0.6$,
$N_{U}=2$, $n_{R}=2$, $n_{U,k}=1$ and 25 dB SNR).}}
\end{figure}

Figure~\ref{fig:graph-Rsum-vs-RF-variousCF} plots the expected sum-rate
$\mathtt{E}[R_{\mathrm{sum}}]$ with respect to the compression output
rate $R_{F}$ for various fronthaul capacity $C_{F}$ with $\epsilon_{F}=0.6$,
$N_{U}=2$, $n_{R}=2$, $n_{U,k}=1$ and 25 dB SNR. Since more packets,
and hence more bits, can be transferred to the cloud within the deadline
$T_{F}$ with increased fronthaul capacity $C_{F}$, the~optimal compression
output rate $R_{F}$ grows with $C_{F}$ for both the MDC and PD~schemes.

\subsection{Optimized Compression Output Rate $R_{F}$}

\begin{figure}[H]
\centering\includegraphics[width=11cm,height=9cm,keepaspectratio]{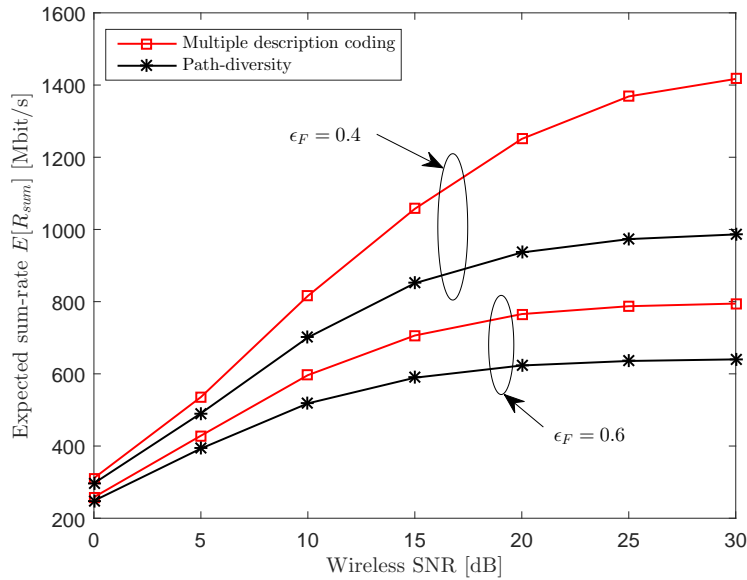}\caption{{\scriptsize{}\label{fig:graph-Rsum-vs-SNR}}{\footnotesize{}Expected
sum-rate $\mathtt{E}[R_{\mathrm{sum}}]$ versus the SNR ($N_{U}=3$,
$n_{R}=3$, $n_{U,k}=1$, $\epsilon_{F}\in\{0.4,0.6\}$ and $C_{F}=100$~Mbit/s).}}
\end{figure}

In this subsection, we present the expected sum-rate $\mathtt{E}[R_{\mathrm{sum}}]$ achieved
when the power allocation variables $\mathbf{P}$, the~quantization noise covariance matrices $\mathbf{\Omega}$ and the compression output rate $R_{F}$ are jointly optimized as
discussed in Section~\ref{sec:Problem-Definition-and}.
In Figure~\ref{fig:graph-Rsum-vs-SNR}, we plot the expected sum-rate
$\mathtt{E}[R_{\mathrm{sum}}]$ versus the SNR for $N_{U}=3$, $n_{R}=3$,
$n_{U,k}=1$, $\epsilon_{F}\in\{0.4,0.6\}$ and $C_{F}=100$ Mbit/s.
We observe from the figure that the MDC scheme shows a larger gain
at a higher SNR level. This suggests that, as~the SNR increases, the~overall performance becomes limited by the quantization distortion
which is smaller for the MDC scheme than for~PD.

\begin{figure}[H]
\centering\includegraphics[width=11cm,height=9cm,keepaspectratio]{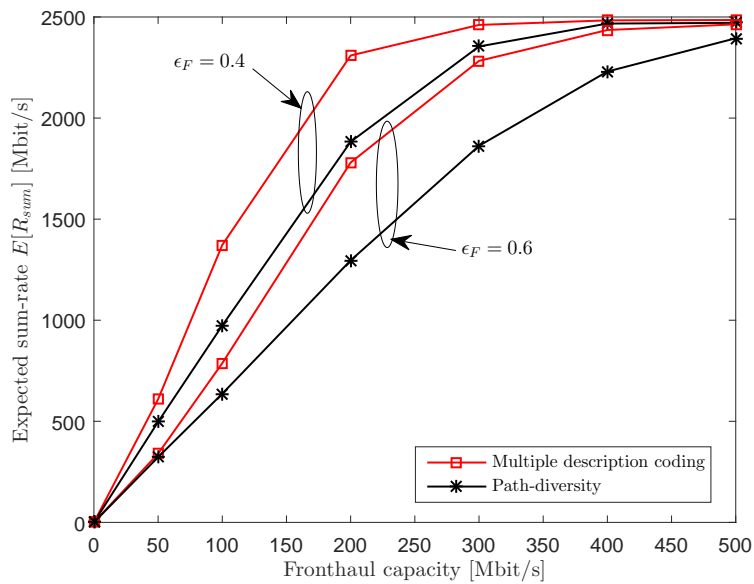}\caption{{\scriptsize{}\label{fig:graph-Rsum-vs-CF}}{\footnotesize{}Expected
sum-rate $\mathtt{E}[R_{\mathrm{sum}}]$ versus the fronthaul capacity
$C_{F}$ ($N_{U}=3$, $n_{R}=3$, $n_{U,k}=1$, $\epsilon_{F}\in\{0.4,0.6\}$
and 25 dB SNR).}}
\end{figure}

In Figure~\ref{fig:graph-Rsum-vs-CF}, we plot the expected sum-rate
$\mathtt{E}[R_{\mathrm{sum}}]$ versus the fronthaul capacity $C_{F}$
for $N_{U}=3$, $n_{R}=3$, $n_{U,k}=1$, $\epsilon_{F}\in\{0.4,0.6\}$
and 25 dB SNR. The~figure illustrates that the MDC scheme shows relevant
gains over the PD scheme in the intermediate regime of $C_{F}$. This
is because, when the fronthaul capacity $C_{F}$ is sufficiently large, the~whole system has a performance bottleneck in the wireless uplink rather
than in the fronthaul network, and~the sum-rate converges to 0 as
$C_{F}$ approaches~0.

\section{Extension to General Numbers of RRHs and Fronthaul~Paths} \label{sec:extension}

In this section, we briefly discuss the application of MDC to the case of general number $N_R$ of RRHs and $N_P$ fronthaul paths.
Each RRH $i$ sends $N_P$ descriptions $\hat{\mathbf{y}}_{i,l}$, $l\in\{1,\ldots,N_P\}$, one on each of the routes to the cloud, where $\hat{\mathbf{y}}_{i,l}$ is a quantized version of the received signal $\mathbf{y}_i$ defined as
\begin{align} \label{eq:quantization-RRH-i-general}
\hat{\mathbf{y}}_{i,l} = \mathbf{y}_{i} + \mathbf{q}_{i,l}.
\end{align}
As in (\ref{eq:quantized-signals-MDC}), under~Gaussian quantization codebook, the~quantization noise $\mathbf{q}_{i,l}$ is independent of $\mathbf{y}_{i}$ and is distributed as $\mathbf{q}_{i,l}\sim\mathcal{CN}(\mathbf{0},\mathbf{\Omega}_i)$.
With MDC, the~cloud can recover the signal $\hat{\mathbf{y}}_{i,l}$ if only the packets for the $l$th description $\hat{\mathbf{y}}_{i,l}$ arrive at the cloud within the deadline. If~a subset of descriptions from RRH $i$ arrive in time, the~cloud can obtain a better signal from RRH $i$, whose quality increases with the size of the subset. Generalizing (\ref{eq:constraint-MDC-1})-(\ref{eq:constraint-MDC-12}), conditions relating the resulting quantization noise covariance matrices and the output compression rate $R_F$ can be found in~\cite{Goyal}.

We define as $M_i\in\{0,1,\ldots,N_P\}$ the number of descriptions of RRH $i$ that arrive at the cloud within the deadline $T_F$. The~probability distribution $p_{M_i}(m) = \Pr[M_i=m]$ of $M_i$ is then given as
\begin{align} \label{eq:probability-number-of-descriptions-general}
p_{M_i}(m) = \sum_{(c_1,\ldots,c_{N_P})\in\{0,1\}^{N_P}} 1\!\left( \sum_{l=1}^{N_P}c_l=m \right) \prod_{l=1}^{N_P} \tilde{P}_l(T_F),
\end{align}
where $1(\cdot)$ is an indicator function that outputs 1 if the input statement is true and 0 otherwise; and the probability $\tilde{P}_l(T_F)$ is defined as $\tilde{P}_l(T_F) = 1(c_l=1) P_l^c(T_F) + 1(c_l=0) (1-P_l^c(T_F))$.

As discussed, with~MDC, the~quality of the information available at the cloud depends on the numbers of descriptions that arrive at the cloud. This means that there are $(N_P+1)^{N_R}$ distinct states depending on the current congestion level of the packet network.
In principle, the~broadcast coding can be applied in such a way that each UE $k$ sends a superposition of $(N_P+1)^{N_R}$ layers. However, this approach does not scale well with respect to $N_R$, and~it is not straightforward to rank all the $(N_P+1)^{N_R}$ states.

To adopt a broadcast coding strategy with a scalable complexity, a~possible option is to fix the number of layers, denoted as $L$, as~in, e.g.,~\cite{Steiner}. Accordingly, the~transmit signal $\mathbf{x}_k$ of each UE $k$ is given by a superposition of $L$ independent signals $\mathbf{x}_{k,l} \sim \mathcal{CN}(\mathbf{0}, P_{k,l}\mathbf{I}_{n_{U,k}})$, $l\in \mathcal{L} = \{1,\ldots, L \}$, i.e.,~$\mathbf{x}_k = \sum_{l\in\mathcal{L}} \mathbf{x}_{k,l}$
with the power constraint $\sum_{l\in\mathcal{L}}P_{k,l} = P$.
We then partition the $(N_P+1)^{N_R}$ congestion  states into $L$ groups, denoted as $\mathcal{S}_1,\ldots,\mathcal{S}_L$, so that the layer-$l$ signals $\{\mathbf{x}_{k,l}\}_{k\in\mathcal{N}_U}$ can be decoded by the cloud for all congestion states in $\mathcal{S}_j$ with $j\geq l$.
Since we can evaluate the probability of all the states using (\ref{eq:probability-number-of-descriptions-general}), the~expected sum-rate can be expressed as a function of the compression output rate,  the~power allocation variables and the quantization covariance matrices. Therefore, we can tackle the problem of jointly optimizing these variables in a similar approach to that proposed in Section~\ref{sec:Problem-Definition-and}.
We leave the evaluation of the impacts of the numbers of RRHs $N_R$ and fronthaul paths $N_P$ to future~work.

\section{Conclusions} \label{sec:conclusion}

In this paper, we have studied the joint design of uplink radio and fronthaul packet transmission strategies for the uplink of C-RAN with a packet-based fronthaul network.
To efficiently use multiple fronthaul paths that carry fronthaul packets from RRHs to cloud, we have proposed an MDC scheme that operates directly on the baseband signals.
Since the signal quality available at the cloud depends on the current network congestion level, a~broadcast coding strategy has been investigated with MDC in order to enable variable-rate transmission.
The advantages of the proposed MDC scheme compared to the traditional PD technique have been validated through extensive numerical results.
Among open problems, we mention the analysis in the presence of imperfect channel state information~\cite{Kang-et-al:TWC}, the~impact of joint decompression of the signals received from multiple RRHs at the cloud~\cite{Park:TVT13,Park-et-al:SPL}, and~design of downlink transmission for C-RAN systems with packet-based fronthaul~network.


\end{document}